\documentclass[aps,prl,groupedaddress,superscriptaddress,twocolumn]{revtex4}

\usepackage{graphicx}
\usepackage{amssymb}
\usepackage{enumerate}

\begin{document}

\title{Localization of preformed Cooper-pairs in disordered superconductors}

\author{Benjamin Sac\'{e}p\'{e}}
\altaffiliation{Present address: Institut N\'{e}el, CNRS and Universit\'{e} Joseph Fourier, BP 166, 38042 Grenoble, France}
\affiliation{SPSMS, UMR-E 9001, CEA-INAC/ UJF-Grenoble 1, 17 rue des martyrs, 38054 GRENOBLE cedex 9, France}
\affiliation{Department of Condensed Matter Physics, The Weizmann Institute of Science, Rehovot 76100, Israel}
\author{Thomas Dubouchet}
\affiliation{SPSMS, UMR-E 9001, CEA-INAC/ UJF-Grenoble 1, 17 rue des martyrs, 38054 GRENOBLE cedex 9, France}
\author{Claude Chapelier}
\affiliation{SPSMS, UMR-E 9001, CEA-INAC/ UJF-Grenoble 1, 17 rue des martyrs, 38054 GRENOBLE cedex 9, France}
\author{Marc Sanquer}
\affiliation{SPSMS, UMR-E 9001, CEA-INAC/ UJF-Grenoble 1, 17 rue des martyrs, 38054 GRENOBLE cedex 9, France}
\author{Maoz Ovadia}
\affiliation{Department of Condensed Matter Physics, The Weizmann Institute of Science, Rehovot 76100, Israel}
\author{Dan Shahar}
\affiliation{Department of Condensed Matter Physics, The Weizmann Institute of Science, Rehovot 76100, Israel}
\author{Mikhail Feigel'man}
\affiliation{L. D. Landau Institute for Theoretical Physics, Kosygin str.2, Moscow 119334, Russia}
\author{Lev Ioffe}
\affiliation{Serin Physics laboratory, Department of Physics and Astronomy, Rutgers University, Piscataway, NJ 08854, USA}

\begin{abstract}

{\bf The most profound effect of disorder on electronic systems is the localization of the electrons transforming an otherwise metallic system into an insulator. If the metal is also a superconductor then, at low temperatures, disorder can induce a dramatic transition from a superconducting into an insulating state. An outstanding question is whether the route to insulating behavior proceeds via the direct localization of Cooper pairs or, alternatively, by a two-step process in which the Cooper pairing is first destroyed followed by the standard localization of single electrons. Here we address this question by studying the local superconducting gap of a highly disordered, amorphous, superconductor by means of scanning tunneling spectroscopy. Our measurements reveal that, in the vicinity of the superconductor-insulator transition, the coherence peaks in the one-particle density of states disappear while the superconducting gap remains intact indicating the presence of localized Cooper pairs. Our results provide the first direct evidence that the transition in our system is driven by Cooper pair localization.}

\end{abstract}

\maketitle

Although superconductivity and Anderson localization lead to the opposite extremes of conductivity at low temperature ($T$), both are due to delicate quantum effects. In superconductors, electrons are bound in Cooper pairs that condense into a zero-resistance, macroscopic, quantum state. In contrast, disorder induces quantum localization of the electron's wave-function that transforms a metal into an insulator with diverging resistance.

It turns out that an increasing level of disorder can cause a transition from a superconductor into an insulator. Understanding how the disorder drives this transition is important for many quantum systems such as amorphous superconductors~\cite{Goldman98}, superconducting nanowires~\cite{Bezryadin00}, high critical-temperatures superconductors~\cite{Steiner05} and ultra-cold atomic gases~\cite{Sanchez10}. Furthermore, this transition is regarded as one of the prototypical quantum phase-transitions driven by disorder in a many-body system, a subject that acquired significant theoretical attention recently. \cite{Basko06,Oganesyan07,Gornyi05}.

For a moderate level of disorder the Anderson theorem~\cite{Anderson Theoreme, Abrikosov}, which is based on the Bardeen-Cooper-Schrieffer (BCS) theory  of superconductivity~\cite{BCS}, states that the critical temperature of superconductivity, $T_c$, remains unchanged. In order to affect $T_c$ significantly a much stronger disorder, at a level that usually causes localization of the electronic wave functions, is needed. As a result the suppression of superconductivity is theoretically expected to be accompanied by a transition to an insulating state~\cite{Kapitulnik1,MaLee85,Kapitulnik2,Sadovskii97,Ghosal98,Ghosal01,Feigelman07}.

The route to the total destruction of superconductivity by disorder can follow two distinct paths. In the first, and more obvious, path disorder-enhanced Coulomb repulsion eliminates Cooper-pairing before the onset of localization. The ensuing poor metal becomes an insulator upon a further increase of disorder~\cite{Finkelstein}. In this case one expects to find an insulating regime similar to that found in other, non-superconducting, metals driven to become insulators by strong disorder.

The second, and more intriguing, path is where the superconductor itself undergoes the transition into an insulating state with only minimal suppression of Cooper-pairing. For such materials,~\cite{Haviland89,Hebard90,Shahar92,Gantmakher98,Steiner05_2,Crane07,Hadacek04,Baturina07_2}  there are two main theoretical mechanisms that have been suggested to explain the transition~\cite{Dubi07}. These mechanisms differ by their emphasis on the relative importance of processes on short and long scales. In non-granular materials, which are the focus of this paper, the short length scale is set by the coherence length of the superconductor. The first of these mechanisms attributes the superconductor-insulator transition (SIT) to the divergence of phase fluctuations at large scales~\cite{Fisher90}. The predictions of this theory are in agreement with resistivity measurements in quench-condensed Bi films~\cite{Goldman98}. The second suggested mechanism~\cite{Feigelman07,Feigelman10} emphasizes the role of the fractal nature of the electron wave-functions at short length scales. Extending earlier works~\cite{Kapitulnik1,MaLee85,Kapitulnik2,Sadovskii97}, this theory predicts that superconductivity at high disorder is maintained by a fragile coherence between a small set of preformed Cooper-pairs that are characterized by an anomalously large binding energy. Consequently, in the vicinity of the SIT, both the insulator and the superconductor are composed of these preformed Cooper-pairs that either localize, leading to an insulating state, or condense into a coherent zero-resistance state.

The existence of preformed Cooper-pairs has been previously inferred from transport measurements findings such as the giant magneto-resistance peak and activated resistance in the magnetic field-tuned SIT of Indium-Oxide (InO)~\cite{Hebard90,Gantmakher98,Shahar04,Steiner05_2} or thin Titanium-Nitride (TiN) films~\cite{Baturina07} and, more recently, from magneto-resistance oscillations with half flux-quantum periodicity in insulating Bi films patterned with a honeycomb array of holes~\cite{Valles07,Valles09}. While these observations are in agreement with the presence of preformed Cooper-pairs at the SIT~\cite{Feigelman10} they only constitute indirect evidence for their existence.
\begin{figure*}
\includegraphics[width=0.7\linewidth]{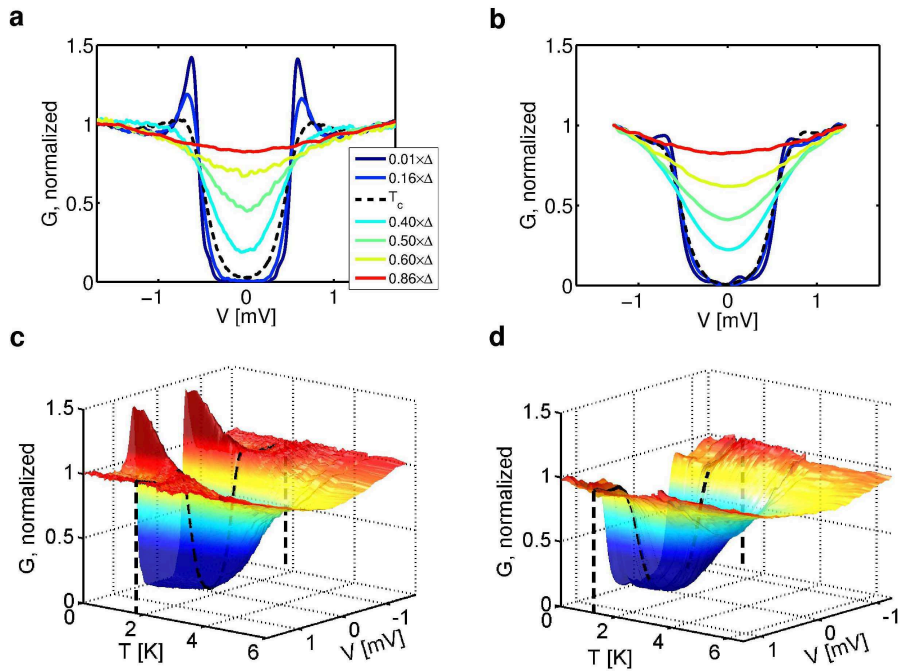}
\caption{{\bf Coherent versus incoherent Cooper pairing revealed by local tunneling spectroscopy.} $T$-evolutions of the local tunneling tunneling conductance $G$ characterized by the presence (\textbf{a},\textbf{c}) or absence (\textbf{b},\textbf{d}) of superconducting coherent peaks. The black dashed lines show the spectra measured at $T_{c}$. \textbf{a},\textbf{b}, Set of spectra for selected temperatures equal to a fraction of the low-T spectral gap. The spectral gap values are $\Delta = 560\, \mu eV$ and $\Delta = 500\, \mu eV$ for \textbf{a} and \textbf{b} respectively (see Methods). \textbf{c},\textbf{d}, Three dimensional view of the same data versus temperature and bias voltage. Spectra of \textbf{a},\textbf{c} and \textbf{b},\textbf{d} were measured on low and high disorder samples respectively.}
\label{FigPseudogap}
\end{figure*}

The existence of Cooper-pairs at short length-scales is more directly revealed by the observation of a superconducting gap in the density of states (DoS)~\cite{BCS,Giaever60} using scanning tunneling microscope (STM) spectroscopy~\cite{Fischer07}. In good BCS superconductors global coherence sets in at $T_c$. Below $T_c$, a DoS suppression begins below the gap energy, $\Delta$, and the lost spectral weight appears as distinct peaks at $\Delta$. These 'coherence' peaks are closely linked to the emergence of a long-range superconducting state. In the case of preformed Cooper-pairs without global coherence, it is theoretically expected~\cite{Feigelman07,Feigelman10} that the spectrum will remain gapped but the coherence peaks will be missing. Only when global superconductivity sets-in, the coherence peaks re-emerge. The height of these coherence peaks is predicted to fluctuate strongly from one location to another.

In this article, we report on a systematic spectroscopic study of the local properties of superconductivity in highly disordered amorphous InO films close to the SIT. Our results, obtained using an STM mounted in a custom-made dilution refrigerator capable of a base temperature of 50 mK, provide the first direct evidence for the existence of preformed Cooper-pairs along with their localization.

\vspace{1cm}

\textbf{Localization of preformed Cooper pairs}

The main feature of our results is the existence of two qualitatively different $T$-evolutions of the DoS spectra as illustrated in Fig.~\ref{FigPseudogap}a and ~\ref{FigPseudogap}b. For $T>T_{c}$ ($T_{c}$ is indicated by the black dashed trace in the figures), both $T$-evolutions exhibit very similar behavior with a low-energy DoS suppression that deepens as $T$ is lowered towards $T_{c}$. A similar DoS suppression above $T_c$ has been seen in other superconductors and has been dubbed the pseudogap~\cite{Timusk99,Fischer07,SacepePseudogapTiN}.

As $T$ is lowered further, the two sets of spectra evolve in a distinctly different fashion. As coherence sets in at $T_{c}$, the spectrum of Fig.~\ref{FigPseudogap}a develops the familiar BCS coherence peaks at $\Delta$. As $T$ is reduced below $T_{c}$ the coherence peaks increase in size, while the DoS at $E<\Delta$ is suppressed further and seems to vanish as $T\rightarrow 0$. In contrast, although a full gap does develop in the spectrum of Fig.~\ref{FigPseudogap}b, with a similar magnitude and a vanishing DoS at low energy, the accompanying coherence peaks are conspicuously absent. The contrast (below $T_c$) and similarities (above $T_c$) between the two types of spectra are highlighted by the corresponding plots of Fig. \ref{FigPseudogap}c and  \ref{FigPseudogap}d, where more complete $T$-evolutions of the spectra are displayed.

The similarity between the $T$-evolution (above $T_c$), as well as the gap magnitude, of both types of spectra indicate that they share the same physical origin. Because the unique shape of the BCS tunneling spectra are known to be the result of the Cooper attraction between electrons, we are led to conclude that the same must be true for the locations where the coherence peaks at the gap edges are missing from the spectra.

The appearance of two types of spectra, similar to those observed in this work, was reported in the numerical study of the two-dimensional attractive Hubbard model with on-site disorder by Ghosal \textit{et al.}~\cite{Ghosal98, Ghosal01}. Qualitatively, they found that the regions where coherence peaks are absent are characterized by rapid fluctuations of the potential that strongly localize the electron states. As a result, they concluded that some of the Cooper-pairs that are formed there are unable to participate in the macroscopic coherent state comprising the condensate.

To develop a quantitative understanding of superconductivity in this high-disorder limit we turn to more recent analytic calculations that take into account the fractal nature of the electronic wave functions close to the localization threshold~\cite{Feigelman07,Feigelman10}. In this theory a pair of electrons occupying a localized state $j$ experience two effects: their net mutual attraction and coherent pair-hopping to neighboring states. The observed spectral gap $\Delta^j$ at a given location contains two distinct contributions : $\Delta^{j}=$ $E_{j}(T) + \Delta _{P}^{j}$. The first term is a BCS-like excitation energy, determined by $E_{j}(T)=(\epsilon _{j}^{2}+h_{j}^{2}(T))^{1/2}$ where $\epsilon _{j}$ is the single electron state energy of the $j$-th state and $h_{j}(T)$ is the local pairing field in this problem. In conventional BCS superconductors the pairing field is uniform and coincides with the single particle gap. In contrast to $E_{j}(T)$, the local contribution to the gap $\Delta _{P}^{j}$ is not related to the development of a global superconducting order-parameter. It results from the Cooper attraction between two electrons populating the same localized state and it is inversely proportional to the volume of that state~\cite{Ghosal01,Feigelman07,Feigelman10}. Because of the fractality of the wave functions the values of $\Delta _{P}^{j}$ are large and fluctuate strongly for nearly-critical wavefunctions $\psi_{j}(\mathbf{r})$. An immediate conclusion of this theoretical analysis~\cite{Feigelman10,IoffeMezard09} is that a global superconducting state survives rather deep into the localized band, up to the region where an average $\overline{\Delta _{P}}$ strongly exceeds typical $h_{j}(T)$. In this regime the local pairing field becomes extremely inhomogeneous and is characterized by a very broad distribution function.

The theoretical framework~\cite{Feigelman10,IoffeMezard09} explains the $T$-evolutions of the spectra displayed in Fig. \ref{FigPseudogap}. For $T>T_c$, global superconductivity is absent ($h_{j}=0$) whereas $\Delta _{P}$ remains non-zero indicating the presence of preformed Cooper-pairs. This regime is revealed by the pseudogap in the DoS shown in Fig.~\ref{FigPseudogap}. For $T<T_c$, the development of superconducting correlations at the location of Fig.~\ref{FigPseudogap}a,c locally gives a non-zero $h_{j}$, as revealed by the progressive growth of the coherence peaks. On the contrary, $h_{j}$ remains nearly zero at the position of Fig.~\ref{FigPseudogap}b,d. The absence, down to our lowest $T$, of coherence peaks in the gapped DoS at some locations is the fingerprint of preformed Cooper pairs that remain localized by the strong disorder and do not participate in the condensate.

\vspace{1cm}

{\bf Proliferation of localized Cooper pairs when approaching the SIT}

\begin{figure}[h!]
\includegraphics[width=1\linewidth]{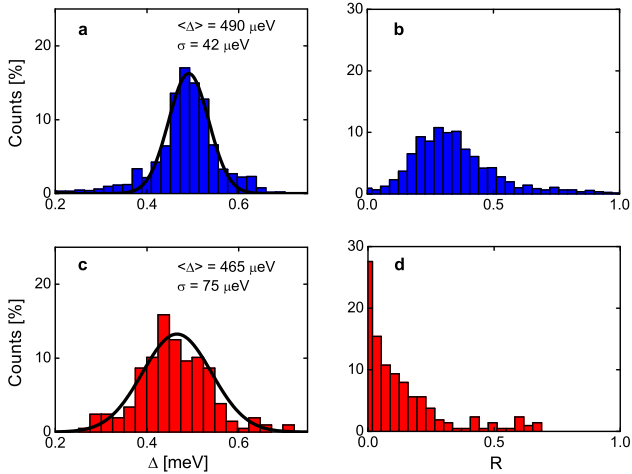}
\caption{{\bf Disorder-induced inhomogeneity and phase incoherence.} Statistics of the local gap value $\Delta $, (\textbf{a,c}), and the local peak-height $R=(G_{peak}-G_{min})/G_{min}$, (\textbf{b,d}) (for definition see text). Blue (red) histograms correspond to the low (high) disorder sample. The difference in the disorder level between the two samples is evidenced by the different $T_c$ ($1.7$ K for low disorder sample and $1.2$ K for high disorder sample) and normal state resistivity (see Supplementary Fig. S1).}
\label{FigStat}
\end{figure}

We now describe the results of a systematic study of our samples. In a disordered system such as ours, a quantitative description is provided by the distribution functions of the main spectra characteristics obtained at different locations on our samples. These characteristics are the gap width at low temperatures (defined in methods section) and the coherence-peak height defined by the ratio $R=(G_{peak}-G_{min})/G_{min}$. Here $G_{peak}$ is the tunneling conductance at the peak energy and $G_{min}$ is the minimum tunneling conductance for an energy just above the gap. In order to study the dependence on disorder we have collected systematic data from high and low disorder samples with $T_c=1.2$ K and $1.7$ K respectively. The results, plotted in Fig.~\ref{FigStat}, represent 208 $I-V$ traces from the high-disorder sample and 2400 traces from the low-disorder sample, each trace measured at a different location.

We begin by inspecting the distribution of $\Delta$. For the low disorder sample (Fig.~\ref{FigStat}a), $\Delta$ is normally distributed (black solid line in the figure) with an average $\overline{\Delta }=490\mu eV$ and standard deviation $\sigma =42\mu eV$. The increase in disorder (Fig.~\ref{FigStat}c) leads to a somewhat lower $\overline{\Delta }=465$ $\mu $eV but a larger $\sigma =75\mu eV$. This broadening of the gap distribution with disorder is similar to that observed previously in TiN films~\cite{Sacepe08}.

A central feature of our results is that the smooth evolution of the gap distribution with disorder is accompanied by a very sharp, qualitative, change of the coherence-peak height-distribution, $\mathcal{P}(R)$, as shown in Figs.~\ref{FigStat}b and \ref{FigStat}d. While the less disordered sample shows a well-defined maximum in the $\mathcal{P}(R)$ distribution function at $R\approx 0.3$, in the more disordered sample, both the maximum of $\mathcal{P}(R)$ and most of its weight are shifted towards $R=0$. This demonstrates that with increasing level of disorder, coherent spectra are replaced by those in which coherent peaks are missing. Close to the SIT, this results in a very inhomogeneous state with superconductivity occupying a small fraction of the sample. The remaining part of the sample is insulating (although the gap in its DoS is also due to Cooper pairing) and its relative area grows with increasing disorder.

Importantly, the fact that incoherently-gapped regions proliferate upon the increase of disorder (see Fig.~\ref{FigStat}) implies that the absence of coherence peaks cannot be a result of surface contamination as there is no reason to expect an abrupt increase of contamination level as the SIT is approached.

\begin{figure}[h!]
\includegraphics[width=0.9\linewidth]{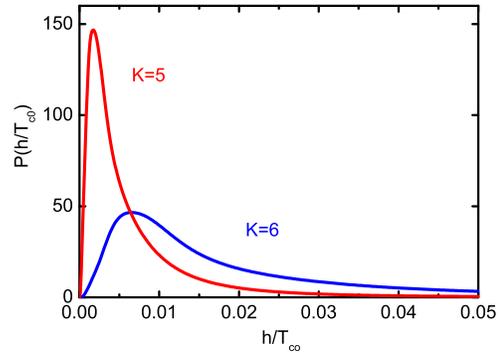}
\caption{{\bf Local pairing field.} Normalized probability distribution of local pairing amplitudes $P(h/T_{c0})$ computed in our model for slightly different disorders characterized by $K=5$ and $K=6$ (see Supplementary Information). The parameters were chosen so that the transition temperature becomes $T_{c0} = 10^{-3} E_F$ in the limit of very low disorder and changes only by a factor 0.9 between $K=6$ and $K=5$. We observe that in contrast to the moderate effect on $T_c$, the distribution functions differ dramatically between these two cases.}
\label{FigLocField}
\end{figure}

Theoretically, the broad statistical distribution of the parameters in our samples is a direct consequence of the fractality of the wave functions, expected to lead to a large and strongly fluctuating values of $\Delta _{P}^{j}$ for nearly-critical wavefunctions $\psi_{j}(\mathbf{r})$~\cite{Feigelman10}. Further, the difference in the distribution functions displayed in Fig.~\ref{FigStat}a,c and \ref{FigStat}b,d is due to the fact that gap magnitudes are controlled by $\Delta _{P}^{j}$ while peak heights are determined by $h_{j}$. To compare to the theoretical predictions we show, in Fig.~\ref{FigLocField}, the computed distribution functions for two levels of disorder. Far from the SIT, the distribution function of $h_{j}$ is relatively narrow and it gradually broadens upon increasing the disorder. In that regime, the typical local field $h$ is roughly equal to the average one as shown by the blue curve in Fig.~\ref{FigLocField}. Closer to the SIT, the distribution function changes dramatically: it acquires a power-law shape $P(h)\sim h^{-\alpha }$ (see red curve in Fig.~\ref{FigLocField}) with exponent $\alpha $ decreasing upon the growth of disorder (see Supplementary Information). The experimentally observed peak height distributions shown in Fig.~\ref{FigStat}b and d conform well to the former and latter situations.
\begin{figure*}[ht!]
\includegraphics[width=0.8\linewidth]{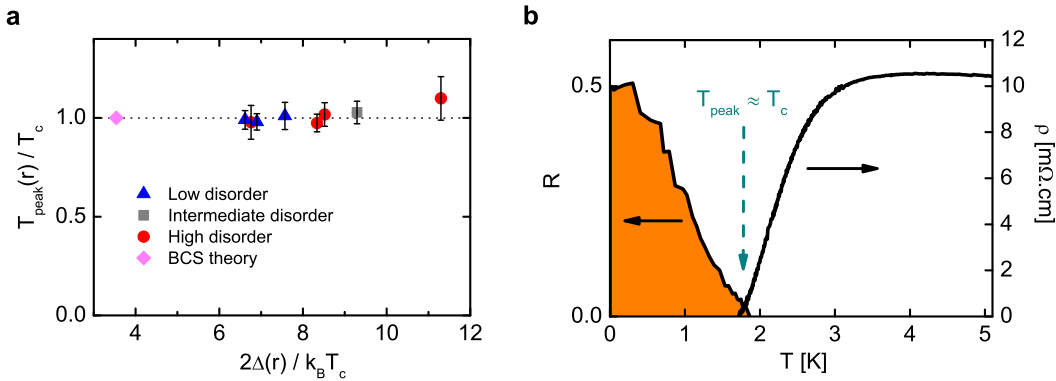}
\caption{{\bf Onset of the superconducting phase coherence.} \textbf{a}, Local onset temperature of the coherence peaks, $T_{peak}(r)$, normalized to $T_c$, versus the ratio $2\Delta (r)/ k_B T_c$ where $\Delta (r)$ is the low temperature spectral gap. For comparison, we added the point $T_{peak}(\mathbf{r})/T_{c}=1$ corresponding to the theoretical BCS ratio $2\Delta /k_{B}T_{c}=3.52$. \textbf{b}, Thermal evolution of the coherence peak height, $R$ (for definition see text), extracted from data of Fig. \ref{FigPseudogap}c and of the resistivity $\rho $ of low disorder sample. This plot evidences the coincidence between the appearance of the zero-resistance superconducting state at $T_c$ with macroscopic phase coherence and the onset of the coherence peaks at $T_{peak}$. }
\label{FigSpectroTransport}
\end{figure*}

Despite the inhomogeneous nature of superconductivity in our samples, we find that nearly all the spectra are characterized by anomalously large $2\Delta /T_{c}$ (see Fig.~\ref{FigSpectroTransport}a). The deviation from the classical BCS value can be understood by noting that, in tunneling experiments, the quantity that is measured is the single-particle DoS and the inferred $\Delta$ is the minimal energy for a single electron, or hole, excitation. In strongly disordered superconductors $\Delta$ is expected to be much larger than the pairing field, as observed by Ghosal et al~\cite{Ghosal98,Ghosal01} in their numerical simulations. One therefore expects that, for increasing disorder, $h_{j}$ will progressively drop to zero, leading to a vanishing $T_{c}$ when approaching the SIT while $\Delta$, whose main part is of \textquotedblleft incoherent\textquotedblright\ origin, will remain finite and may even increase. In other words, a local spectral gap due to Cooper attraction can exist without global superconductivity leading in a natural way to the large $2\Delta /k_{B}T_{c}$ ($k_B $ is Boltzmann constant).

\vspace{1cm}

{\bf Macroscopic quantum phase coherence probed at a local scale}

Finally, to clarify how phase coherence is established in such an inhomogeneous superconducting state, we have studied the $T$ evolution of several spectra measured at different locations \textbf{r} and on samples with different $T_{c}$. For each evolution, we plot in Fig.~\ref{FigSpectroTransport}a $T_{peak}(\mathbf{r})/T_{c}$, where $T_{peak}(\mathbf{r})$ is the temperature below which coherence peaks start to grow (see Fig.~\ref{FigSpectroTransport}b) versus the local $2\Delta (\mathbf{r})/k_{B}T_{c}$. While this latter ratio can vary between 6.5 and 11.5, $T_{peak}(\mathbf{r})/T_{c}$ remains nearly 1. This proves that peaks appear when and only when global superconductivity is established regardless of the local gap. In this respect, our InO films behave very differently from granular superconductors where one would expect a constant $2\Delta (\mathbf{r})/k_{B}T_{peak}(\mathbf{r})$ and a macroscopic $T_{c}$ determined by the interplay of the charging and Josephson energies in the array of grains~\cite{Efetov}.

The appearance of coherence peaks at the same temperature where resistance is vanishing (see Fig.~\ref{FigSpectroTransport}b) allows us a direct comparison between local and global measurements. This is especially striking since both the gap magnitude and peak heights fluctuate very strongly across our samples, while local coherence peaks appear simultaneously at $T_{c}$. This counter-intuitive behavior allows us to rule out a scenario where preformed Cooper-pairs would condense locally at different temperatures above $T_c$ leading to independent superconducting droplets. For such an inhomogeneous sample, the zero resistance state appearing at $T_c$ would simply correspond to the percolation of these droplets. In a disordered system such as ours, because of their fractal nature, the electronic wave-functions spread far in the sample and therefore weakly overlap with a large number of neighboring states. At $T_c$ these wavefunctions condense into a single superconducting state despite the local fluctuations of the pairing field $h_{j}$. Accordingly, coherence peaks emerge everywhere at $T_c$ but their height vary from place to place.

To conclude, we presented the evidence that electrons remain Cooper paired at the SIT in disordered InO films. The transition is driven by the increase in the number of incoherent pairs at the expense of the ones that participate in the condensate. Close to the transition, on the superconducting side, $\Delta$ fluctuates strongly indicating a spontaneously formed inhomogeneity. The perseverance of single electron gap implies that very close to the transition on the insulating side, the transport is dominated by incoherent Cooper pair hopping. The properties of the material in this regime might be very unusual and deserve further studies.

Note that very recent numerical simulations~\cite{Bouadim10} are consistent with our interpretation of the data.

\section{Methods}

\textbf{Samples.} Our samples are disordered thin films of amorphous indium oxide. The films are prepared by using electron-beam
evaporation of high purity ($99.999\%$) $In_{2}O_{3}$ onto $SiO_{2}$ in an $O_{2}$ background. The thicknesses of the films studied here are 150 and 300 \text{\AA } as measured \textit{in situ} by a quartz crystal thickness monitor. STM measurements give a typical rms surface roughness of 1 nm on a scanned surface of  $1\, \mu m $. Transmission electron microscopy studies revealed the amorphous nature of the samples without detecting any crystalline inclusion~\cite{Shahar92}. In order to perform transport measurements, samples are patterned into Hall bridges via a shadow mask and contacts are made using pressed indium and gold wire.

\textbf{Measurements.} Transport measurements and tunneling spectroscopy were systematically carried out during the same experiment in a home-built STM cooled down to 50mK in an inverted dilution refrigerator. Temperature of the sample holder, which is weakly coupled to the dilution refrigerator, was accurately controlled by a $RuO_{2}$ thermometer and a resistive heater. No measurable thermal drift of the tip position occurred in our experiments between the base temperature of 50mK and the highest measured temperature of about 6 K. To perform spectroscopy, the STM Pt/Ir tip was aligned at the center of the sample Hall bridge. The differential conductance of the tunnel junction, $G(V)=\frac{dI}{dV}$, was then measured by a lock-in amplifier with an alternative voltage modulation of $10-30 \mu \text{V}$ added to the ramped bias voltage thus allowing to probe the local DoS. The energy resolution of the spectroscopy can be described by the effective temperature, $T_{eff}\simeq 0.3\,K$, used to fit the superconducting spectra with the theoretical BCS DoS. The discrepancy between $T_{eff}$ and the sample thermometer is due to unfiltered electromagnetic radiations which heat the electrons and from voltage noise generated by room temperature electronics~\cite{Lesueur06}. The tunneling current was $0.05-1\text{nA}$ for millivolts bias voltage yielding a tunneling resistance in the $\text{M} \Omega $ range, which is much higher than the film resistance ($\lesssim 30\, k\Omega$) between the STM junction and the contact. This ensures a negligible voltage drop across the resistive film in series with the STM junction during spectroscopy above $T_c$. For statistics analysis of the gap distribution we used the practical definition of the gap $\Delta /e = 1.1 V_{max}$, where $e$ is the electron charge and $V_{max}$ was defined as the bias voltage at which the numerically computed $dG(V)/dV $ is maximum at the gap edge. The factor $1.1$ compensates the shift of $V_{max}$ to lower values that is due to thermal broadening of the differential tunneling conductance. This relation which is valid for superconducting spectra was also used to quantify the gap value for spectra without peaks. Transport measurements of the films (see Supplementary Information) were carried out with a low frequency lock-in amplifier technique in a four terminal configuration with excitations currents of $1\text{nA}$ below $5K$ and of $10\text{nA}$ at higher temperatures.

\vspace{1cm}

\textbf{Acknowledgements} We thank A. Finkel'stein, V. Kravtsov, M. M\'{e}zard, Z. Ovadyahu and N. Trivedi for valuable discussions. D.S. and M.O. acknowledge the Israeli Science Foundation and the Minerva Fund.

\vspace{1cm}

\textbf{Author Contributions} B.S. and T.D. equally contributed to this work.



\section{Supplementary Information}

{\bf Transport measurements}
\makeatletter\renewcommand{\thefigure}{S\@arabic\c@figure}

\begin{figure}[h!]
\includegraphics[width=1\linewidth]{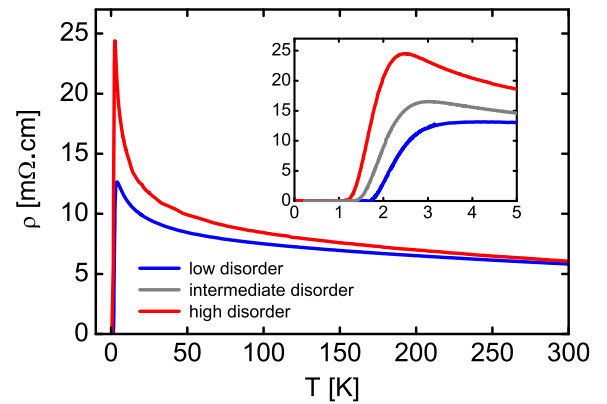}
\caption{{\bf Resistivity versus temperature of amorphous indium oxyde films.} Thicknesses of the samples are 150 \text{\AA } for low disorder sample and 300 \text{\AA } for high and intermediate disorder samples. Inset: superconducting transitions measured during the cooling of the STM setup. The critical temperatures defined as the vanishing resistance (see main text) are $1.7\,K$, $1.4\,K$ and $1.2\,K$ for low, intermediate and high disorder sample respectively.}
\label{FigRT}
\end{figure}

In this section we present the transport properties of our highly disordered amorphous InO films. The $T$-evolutions of the resistivity for the three samples studied in this work are shown in Fig. \ref{FigRT}. All films present an insulating trend: The resistivity increases as $T$ is lowered from room temperature. Despite this clear signature of electron localization, the films undergo at low temperature a superconducting transition with critical temperature $1-2\,K$ (see inset of Fig.~\ref{FigRT}). The continuous decrease of the critical temperature upon increasing disorder is the hallmark of the disorder-driven superconductor-insulator transition that occurs in our InO films.

{\bf Summary of the theory of superconductivity close to the mobility edge}

Here we sketch the main ingredient of the theory developed in~\cite{Feigelman07bis,Feigelman10bis,IoffeMezard09bis,FeigelmanIoffeMezardbis} and its application to the data. In the range of strongly developed pseudogap $\Delta _{P}\gg T_c$ electron \textquotedblleft orbitals\textquotedblright\ $\psi _{j}(\mathbf{r})$ are populated by either zero or two electrons at low $T$'s. In this case the whole Hilbert space of the electronic problem reduces to the \textquotedblleft pseudospin\textquotedblright\ subspace described by operators formally equivalent to spin-$\frac{1}{2}$ variables $S_{j}^{\pm },S_{j}^{z}$ associated with each orbital: $S_{j}^{+}=c_{\uparrow ,j}^{+}c_{\downarrow ,j}^{+}$ is the pair creation operator and $2S_{j}^{z}=c_{\uparrow ,j}^{+}c_{\uparrow ,j}+c_{\downarrow,j}^{+}c_{\downarrow,j}-1$; this representation was proposed originally by Anderson in~\cite{AndersonPseudospins}. The development of superconducting coherence is described by the pseudospin Hamiltonian
\begin{equation}
H_{PS}=2\sum_{j}\epsilon
_{j}S_{j}^{z}-\frac{g}{2}\sum_{ij}M_{ij}(S_{i}^{+}S_{j}^{-}+S_{i}^{-}S_{j}^{+}),
\label{HamSpin}
\end{equation}
with matrix elements $M_{ij}=\int d^{3}\mathbf{r}\psi _{i}^{2}(\mathbf{r})\psi_{j}^{2}(\mathbf{r})$ and coupling constant $g$. On-site energies $\epsilon _{j}$ are distributed over wide band with bare DoS $\nu _{0}$. The pseudogaped regime $\Delta _{P}\gg T_c$ is realized~\cite{Feigelman07bis,Feigelman10bis} when typical level spacing inside localization volume, $\delta \sim 1/\nu _{0}\xi_{loc}^{3}$, is the largest energy scale, $\delta \gg \Delta _{P}$. The local pairing fields $h_{j}(T)$ determine the  average transverse pseudospin component via $\langle S_{i}^{x}\rangle =(h_{i}/2E_{i})\tanh(E_{i}/T)$; they obey self-consistency equation
\begin{equation}
h_{i}=\frac{g}{2}\sum_{j}M_{ij}\frac{h_{j}}{E_{j}}\tanh \frac{E_{j}}{T}\,,\quad
E_{j}=\sqrt{h_{j}^{2}+\epsilon _{j}^{2}.}  \label{h-Eq}
\end{equation}
At very large $\delta $, long-range coherence is not established even at $T=0 $, and insulating ground state takes over. The quantum phase transition between the ordered state with non-zero $\left\langle S^{x,y}\right\rangle\neq 0$ and disordered one with $\left \langle S^{x,y}\right\rangle =0$ was studied in Ref.~\cite{IoffeMezard09bis,FeigelmanIoffeMezardbis} within the simplified model defined by the Hamiltonian (\ref{HamSpin}) on the Bethe lattice with coordination number $K$ and all nonzero couplings $M_{ij}=1/K$. The model is characterized by dimensionless coupling $\lambda =g\nu _{0}\ll 1$. The result of this study is that in the limit $K>K_{1}=\lambda e^{1/2\lambda }$ usual
BCS-like mean-field theory works well and uniform superconducting state occurs below $T_{c0} =\nu _{0}^{-1}e^{-1/\lambda }$. In particular, below $T_{c0}$ nonzero thermal average $h_{j}(T)$ appear.

Fluctuations become important at $K<K_{1}=\lambda e^{1/2\lambda }$. They lead to two effects: critical temperature $T_{c}(K)$ start to drop down with $K$ decrease and the order parameter $h_{i}(T)$ becomes strongly inhomogeneous as function of the site number $i$ and cannot be expressed in terms of local energy $\epsilon _{j}$ \thinspace only. Eventually, transition temperature vanishes at $K=K_{2}=\lambda e^{1/e\lambda }$, the ground-state at $K<K_{2}$ is a many-body insulator, with discrete spectrum of low-lying excitations.

In the whole region $K_{2}<K<K_{1}$ on the superconducting side of quantum phase transition, nonzero order parameter $h_{j}(T)$ appears below well-defined global transition temperature $T_{c}(K)$ and leads to the growth of coherence peaks seen in the tunneling spectra, Fig.~1a,c. At the same time, the relative heights of these peaks are proportional to the values of $h_{j}(T)$ for the electron
orbitals $\psi _{j}(\mathbf{r})$ which have considerable weights near the tip position $\mathbf{r}$. The values of $h_{j}$ fluctuate strongly between different sites $j$ with very close local energies $\epsilon _{j}$. The strength of these fluctuations is characterized by the distribution function $P(h)$ with a long tail: close to the transition line $T_{c}(K)$ the distribution is $P(h)\sim (1/h_{0})(h_{0}/h)^{\alpha}$ in a wide range of $h/h_{0}$. The exponent $\alpha$ decreases with decrease of $K$, i.e. in physical terms, with increase of disorder. In particular, at $K_{2}<K<K_{1}$ the exponent $1<\alpha<2$ was found~\cite{IoffeMezard09bis,FeigelmanIoffeMezardbis}, indicating that simple average $h_{av}=\langle h\rangle $ and typical value $h_{typ}=exp(\langle \log (h)\rangle )$ differ qualitatively, with $h_{typ}\ll h_{av}$\thinspace\ ( $h_{typ}$ characterizes behaviour of \textit{almost any} specific sample, whereas $h_{av}$ corresponds to a contribution of extremely rare fluctuations in the ensemble average over many samples). We show in Fig.~3 theoretical results for two nearby values of the coordination number
$K= 5$ and $K=6$, both at the dimensionless coupling constant $\lambda= 0.128$. While the values of the critical temperature are close in these two cases ($T_c(K=5)/T_c(K=6) \approx 0.9$), the shape of $P(h)$ distribution changes considerably: the "tail" with $ h \ll h_{typ}$ becomes much more pronounced for the $K=5$ case that corresponds to stronger disoder. This theoretical result is in good agreement with our observations shown in Fig.~2b,d: distribution of peak heights is moderately narrow in the less disordered sample shown in Fig.~2b and becomes very broad (like the $P(h)$ distribution function at $K<K_{1}$) for the more disordered sample in Fig.~2d.

\end{document}